\documentclass[a4paper]{article}

\usepackage{INTERSPEECH2021}
\usepackage{multirow}
\title{AdaVocoder: Adaptive Vocoder for Custom Voice}
\name{Xin Yuan, Yongbing Feng, Mingming Ye, Cheng Tuo, Minghang Zhang}
\address{VXI, China}
\email{\{xin.yuan, yongbing.feng, mingming.ye, cheng.tuo, minghang.zhang\}@vxichina.com}

\begin{document}

\maketitle
\begin{abstract}
Custom voice is to construct a personal speech synthesis system by adapting the source speech synthesis model to the target model through the target few recordings. The solution to constructing a custom voice is to combine an adaptive acoustic model with a robust vocoder. However, training a robust vocoder usually requires a multi-speaker dataset, which should include various age groups and various timbres, so that the trained vocoder can be used for unseen speakers. Collecting such a multi-speaker dataset is difficult, and the dataset distribution always has a mismatch with the distribution of the target speaker dataset.

This paper proposes an adaptive vocoder for custom voice from another novel perspective to solve the above problems. The adaptive vocoder mainly uses a cross-domain consistency loss to solve the overfitting problem encountered by the GAN-based neural vocoder in the transfer learning of few-shot scenes. We construct two adaptive vocoders, AdaMelGAN and AdaHiFi-GAN. First, We pre-train the source vocoder model on AISHELL3 and CSMSC datasets, respectively. Then, fine-tune it on the internal dataset VXI-children with few adaptation data. The empirical results show that a high-quality custom voice system can be built by combining a adaptive acoustic model with a adaptive vocoder.

\end{abstract}

\section{Introduction}
As the development of deep learning\cite{lecun2015deep}, neural network-based text to speech (TTS) has thrived\cite{tan2021survey}. Nowadays, since the voice synthesized by the TTS model trained by a large amount of high-quality recordings has been comparable to those of human beings\cite{shen2018natural}\cite{ren2021fastspeech}\cite{elias21_interspeech}\cite{yamamoto2020parallel}\cite{kong2020hifi}, more and more people are interested in custom voice\cite{chen2021adaspeech}. Custom voice has attracted attention in different scenarios such as personal assistants, voice navigation and audiobooks. At the same time, custom voice research has also increased in academia\cite{yan2021adaspeech}\cite{yan2021adaspeech3}. The main problem is the contradiction between the lack of custom recordings and the data-driven property of deep learning at present.

Now the main pipeline of high-quality TTS system consists of three modules, namely TTS front-end, acoustic model and vocoder\cite{tan2021survey}. The TTS front-end mainly includes text analysis which does not affect custom voice. The main influences of custom voice are acoustic model and vocoder. Scholars have conducted research on the two modules respectively.

From the perspective of acoustic model, there are also several distinctive
challenges in custom voice: 1)Custom recordings usually have different acoustic conditions such as prosody and emotion from the source speech data (the data used to train the source TTS model), resulting in poor adaptation. 2) Due to the large acoustic model, it is generally run in the cloud. In order to reduce memory capacity and service cost, a trade-off between fine-tuning parameters and speech quality is required. The most common custom voice acoustic model is to fine-tune the source model\cite{chen2018sample}\cite{kons2019high} or the decoder part of the source model to achieve adaptive results\cite{moss2020boffin}\cite{zhang2020adadurian}, but it will result in lots of adaptation parameters. Some papers only fine-tune the speaker embedding to reduce the adaptive parameters\cite{arik2018neural}, which will lead to poor speech quality. The series of articles about AdaSpeech proposes Conditional Layer Normalization, which uses id information of speaker as a condition for layer normalization\cite{chen2021adaspeech}\cite{yan2021adaspeech}\cite{yan2021adaspeech3}. And only Conditional Layer Normalization is transferred in transfer learning, which ensures the reduction of adaptive parameters and the effect of custom voice. AdaSpeech can create a high-quality custom voice acoustic model using only 10 samples of the custom recordings.

From the vocoder perspective, the main problem is similar to the first problem mentioned in acoustic model problem is the difference between the distribution of custom recordings data and the distribution of source recordings data, which leads to poor adaptation quality\cite{hsu2019towards}. Since the vocoder model is small and generally runs on the terminal side, there is absence of trade-off between fine-tuning parameters and voice quality. Solving the problem of vocoder adaptation is usually to build a robust vocoder\cite{hsu2019towards}\cite{lorenzo2018towards}\cite{jang2020universal}. \cite{lorenzo2018towards} reported that a WaveRNN-based neural vocoder trained on multi-speaker multilingual data can generate natural speech despite conditions from an unseen domain. But the data distribution of the unseen speaker is usually similar to that of the training data set. If the age of the unseen speaker is larger than that of the speaker in the training data set, the quality of the adaptive vocoder will be greatly reduced. Some scholars have designed a large number of experiments to compare the robustness of the more popular vocoders (such as WaveNet\cite{oord2016wavenet}, WaveRNN\cite{kalchbrenner2018efficient}, FFTNet\cite{jin2018fftnet} and Parallel WaveGAN\cite{yamamoto2020parallel}). They found that speaker variety is incredibly important to implement a universal vocoder\cite{hsu2019towards}. However, it is more difficult to collect variety speaker corpora in real life. The difference between the source domain data distribution and the target domain generates that the custom voice cannot reach the Human level. Another solution to the problem of vocoder adaptation is fine-tuning the model. HooliGAN verified that the fine-tuning model can be better conditional on 30 minutes recordings(582 sentences)\cite{mccarthy2020hooligan}. But 30 minutes of recordings is too long and difficult to collect for custom voice. Training neural vocoders, such as HiFi-GAN, on a target domain containing limited examples (e.g., 10) can easily result in overfitting.

This paper utilizes a novel cross-domain distance consistency loss to preserve the relative similarity and differences between instances in the source, thereby preventing overfitting caused by fine-tuning vocoders under limited instances. The contributions of this work are:
\begin{itemize}
\item we propose a adaptive vocoder for custom voice.
\item We analyze the fine-tuning results of the general GAN-based vocoders on few-shot (under 10 examples) scenario based on a mean opinion score (MOS) survey.
\end{itemize}

In Section \ref{Related work}, general GAN-based vocoders and cross-domain correspondence are introduced. In Section \ref{Methods}, we propose our approach to address vocoder adaptation by exploiting a cross-domain distance consistency loss. 
We present the experimental results in Section \ref{Experiments} and introduce the dataset and evaluation criteria. We finally summary our results in Section \ref{Conclusion}.

\section{Related Works}\label{Related work}
\subsection{GAN-Based Vocoders}
Nowadays, neural vocoders have replaced traditional heuristic methods and dramatically enhanced the quality of generated speech\cite{hsu2019towards}. Generative adversarial networks (GANs) have been widely used in data generation tasks, such as image generation\cite{zhu2017unpaired}, text generation\cite{yu2017seqgan}, and audio generation\cite{donahue2018adversarial}. GAN consists of a generator for data generation, and a discriminator to judge the authenticity of the generated data. A lot of vocoders leverage GAN to ensure the audio generation quality, including WaveGAN\cite{donahue2018adversarial}, MelGAN\cite{kumar2019melgan}, Parallel WaveGAN\cite{yamamoto2020parallel}, HiFi-GAN\cite{kong2020hifi}, etc. Most GAN-based vocoders generate waveform sequence from mel-spectrograms, which require an upsampling process. This upsampling process is usually done using dilated convolutions. Identifying long-term dependencies is the key for modeling realistic speech audio. MelGAN\cite{kumar2019melgan} proposed the multi-scale discriminator to capture consecutive patterns and long-term dependencies. HiFi-GAN\cite{kong2020hifi} proposed the multi-period discriminator consisting of several sub-discriminators handling a portion of periodic signals of input audio respectively. HiFi-GAN can efficiently synthesize high quality speech audio. This paper experimentes with fine-tuning Mel-GAN and HiFi-GAN.

\subsection{Cross-Domain Correspondence}
The primary problem of this paper is to obtain an adaptive vocoder with a small number of samples. Typically, training is performed on a large dataset (source dataset) and then fine-tuned with a small number of samples (target dataset). Domain adaptation has emerged as a new learning technique to address the lack of massive amounts of labeled data\cite{wang2018deep}. For instance, the domain-adversarial neural network (DANN)\cite{ganin2015unsupervised} integrates a gradient reversal layer (GRL) into the standard
architecture to ensure that the feature distributions of the source and target domains are similar. However, domain adaptation is generally applied to classification problems, including learning a feature similarity function between the query and support examples\cite{vinyals2016matching}\cite{snell2017prototypical} and learning how to adapt a base-learner to a new task\cite{finn2017model}\cite{nichol2018first}. Cross-domain correspondence is to solve the problem of overfitting encountered in few-shot learning in generative models. For example, DistanceGAN\cite{benaim2017one} proposes to preserve the distance between input pairs in the corresponding generated output pairs, thereby reducing the problem of overfitting. \cite{ojha2021few} proposed to inherit the learned diversity from the source model to the target model through the cross-domain distance consistency loss, thereby solving the problem of easy overfitting in few-shot transfer learning. This paper draws on the cross-domain consistency loss and adds this loss to the GAN-based vocoders to solve the overfitting problem encountered by the few-shot adaptive vocoders.

\section{Methods}\label{Methods}
We get a source generator $G_s$, trained on a large source dataset $D_s$, which maps mel-spectrograms $m\sim p_m(m) \subset M$ into waveforms $x$. AdaVocoder aims to learn an adapted generator $G_{s \to t}$ by initializing the weights to the source generator and fitting it to small target dataste $D_t$.
A traditional fine-tuning method is to directly use the trained generator and discriminator, and then use the GAN training step for fine-tuning. The solution formula is as follows:

\begin{equation}
\begin{aligned}
& \mathcal{L}_{\mathrm{adv}}(G, D)=D(G(m))-D(x) \\
G_{s \rightarrow t}^{*}=& \mathbb{E}_{m \sim p_{m}(m), x \sim \mathcal{D}_{t}} \arg \min _{G} \max _{D} \mathcal{L}_{\mathrm{adv}}(G, D).
\end{aligned}
\end{equation}

\cite{wang2018transferring} shows that the above transfer method performs better when the target dataset size is larger than 1000 samples. However, in few-shot transfer learning, the discriminator can remember these few samples and force the generator to reconstruct these samples, which leads to overfitting. AdaVocoder utilizes a cross-domain consistency loss to address this overfitting phenomenon.

\begin{figure}[t]
  \centering
  \includegraphics[width=\linewidth]{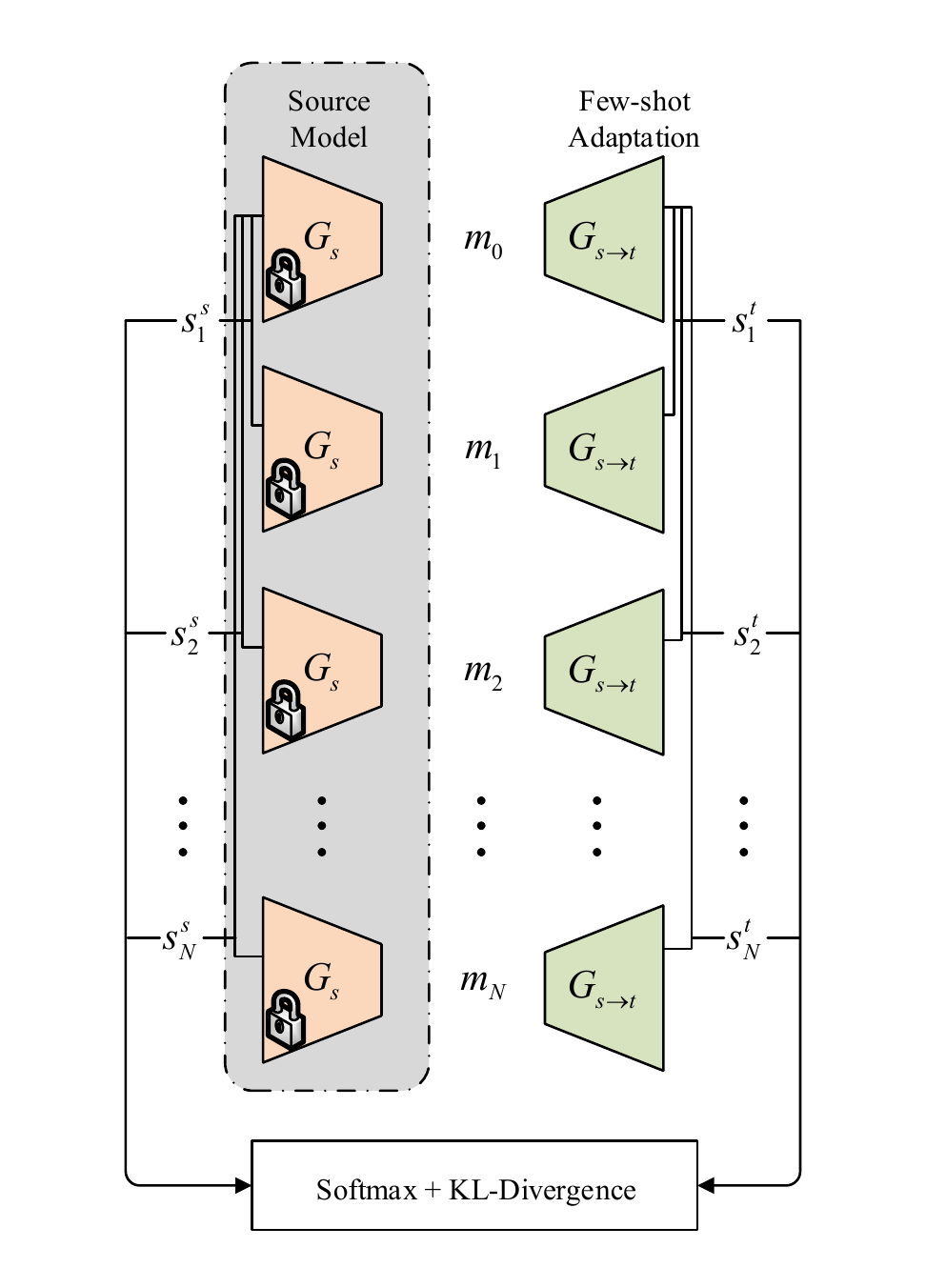}
  \caption{Cross-domain consistency loss}
  \label{fig:interspeech}
\end{figure}

\subsection{Cross-domain Distance Consistency}
Cross-domain distance consistency\cite{ojha2021few} was first proposed to solve the problem of diversity in image generation. The correspondence to speech is similar during the training process of GAN. We enforce pair-wise relative distances to prevent overfitting. 

The specific construction process is as follows:
we sample a batch of $N+1$  mel-spectrograms $\left\{m_{n}\right\}_{0}^{N}$ and use their pairwise similarities in feature space to construct $N$-way probability distributions for each voice. This is illustrated in Fig.\ref{fig:interspeech} from the viewpoint of $m_0$.

The probability distributions of source and adapted generators for the $i^{th}$ mel-spectrogram are given by, 

\begin{equation}
\begin{aligned}
y_{i}^{s, l} &=\operatorname{Softmax}\left(\left\{\operatorname{sim}\left(G_{s}^{l}\left(m_{i}\right), G_{s}^{l}\left(m_{j}\right)\right)\right\}_{\forall j \neq i}\right) \\
y_{i}^{s \rightarrow t, l} &=\operatorname{Softmax}\left(\left\{\operatorname{sim}\left(G_{s \rightarrow t}^{l}\left(m_{i}\right), G_{s \rightarrow t}^{l}\left(m_{j}\right)\right)\right\}_{\forall j \neq i}\right),
\end{aligned}
\end{equation}
where sim denotes the cosine similarity between generator activations at the $l^{th}$ layer.
The cross-domain consistency loss is:
\begin{equation}
\mathcal{L}_{\mathrm{dist}}\left(G_{s \rightarrow t}, G_{s}\right)=\mathbb{E}_{\left\{m_{i} \sim p_{m}(m)\right\}} \sum_{l, i} D_{K L}\left(y_{i}^{s \rightarrow t, l} \| y_{i}^{s, l}\right).
\end{equation}

\subsection{AdaMelGAN}
The generator and discriminators of AdaMelGAN are the same as MelGAN, except that a cross-domain consistency loss is added during the training process. Next, we introduce the generator and discriminators of MelGAN, and propose how to introduce cross-domain consistency loss into MelGAN.

\subsubsection{Generator}
The generator of MelGAN is a fully convolutional feed-forward network with mel-spectrogram $m$ as input and raw waveform $x$ as output. Since the mel-spectrogram is a 256× lower temporal resolution, MelGAN use a stack of transposed convolutional layers to upsample the input sequence. Details can be found in reference \cite{kumar2019melgan}.

\subsubsection{Discriminator}
The discriminators of MelGAN focus on how to design models to capture the characteristics of waveform, in order to provide better guiding signal for the generator.
MelGAN adopts a multi-scale architecture with 3 discriminators ($D1$,$D2$,$D3$) that have identical network structure but operate on different audio scales. $D1$ operates on the scale of raw audio, whereas $D2$ and $D3$ operate on raw audio downsampled by a factor of 2 and 4 respectively.

\subsubsection{Final Objective}
To train MelGAN on the source data, we use the same approach as the original paper, i.e. use the hinge loss version of the GAN objective.
\begin{equation}
\begin{aligned}
    \mathcal{L}_{\mathrm{adv}}(D_k, G)= &\mathbb{E}_{x}\left[ \min \left(0,1-D_{k}(x)\right)\right] + \\ &\mathbb{E}_{m}\left[\min \left(0,1+D_{k}(G(m))\right)\right], \forall k=1,2,3 \\
    \mathcal{L}_{\mathrm{adv}}(G, D)= &\mathbb{E}_{m}\left[\sum_{k=1,2,3}-D_{k}(G(m))\right]
\end{aligned}
\end{equation}

In addition, feature matching loss ($
\mathcal{L}_{\mathrm{fm}}\left(G, D\right)
$)\cite{kumar2019melgan} is introduced to train generators. 

In few-shot transfer learning, we introduce the cross-domain consistency loss into the training process of the generator of MelGAN.
\begin{equation}
\begin{aligned}
G_{s \rightarrow t}^{*}=\arg \min _{G_{s \rightarrow t}} \max _{D} & \mathcal{L}_{\text {adv }}\left(G_{s \rightarrow t}, D\right) +\lambda_{cd} \mathcal{L}_{\text {dist }}\left(G_{s \rightarrow t}, G_{s}\right)  \\ & +\lambda_{fm} \sum_{k=1}^{3} \mathcal{L}_{\mathrm{fm}}\left(G_{s \rightarrow t}, D_{k}\right),
\end{aligned}
\end{equation}
where, $\lambda_{cd}$ is the coefficient of cross-domain consistency loss, and $\lambda_{fm}$ denotes the coefficient of feature matching Loss, which are $10^3$ and 10 respectively.

\subsection{AdaHiFi-GAN}
The generator and discriminators of AdaHiFi-GAN are the same as HiFi-GAN, except that a cross-domain consistency loss is added during the training process. Next, we introduce the generator and discriminators of HiFi-GAN, and propose how to train the AdaHiFi-GAN.

\subsubsection{Generator}
Similarly to MelGAN, HiFi-GAN also uses a fully convolutional neural network to build a generator. However, HiFi-GAN processes different patterns of various lengths in parallel through a multi-receptive field fusion module.

\subsubsection{Discriminator}
HiFi-GAN contains two types of discriminators, namely multi-period discriminator (MPD) and multi-scale discriminator (MSD, proposed in MelGAN). MPD can capture different implicit structures by looking at different parts of an input audio in different periods.

\subsubsection{Final Objective}
HiFi-GAN losses for the generator $G$ and the discriminator $D$
are defined as
\begin{equation}
\begin{aligned}
&\mathcal{L}_{\mathrm{adv}}(D, G)=\mathbb{E}_{(x, m)}\left[(D(x)-1)^{2}+(D(G(m)))^{2}\right] \\
&\mathcal{L}_{\mathrm{adv}}(G, D)=\mathbb{E}_{m}\left[(D(G(m))-1)^{2}\right].
\end{aligned}
\end{equation}
In addition to the optimization objective of GAN, HiFi-GAN adds a mel-spectrogram loss ($
\mathcal{L}_{\mathrm{mel}}\left(G\right)
$) and feature matching loss ($
\mathcal{L}_{\mathrm{fm}}\left(G, D\right)
$) to improve the training efficiency of the generator and the fidelity of the generated audio. 

In few-shot transfer learning, we introduce the cross-domain consistency loss into the training process of the generator of HiFi-GAN.
\begin{equation}
\begin{aligned}
G_{s \rightarrow t}^{*}=\arg \min _{G_{s \rightarrow t}}& \max _{D}  \mathcal{L}_{\text {adv }}\left(G_{s \rightarrow t}, D\right) +\lambda_{cd} \mathcal{L}_{\text {dist }}\left(G_{s \rightarrow t}, G_{s}\right)  \\ & +\lambda_{fm} \mathcal{L}_{\mathrm{fm}}\left(G_{s \rightarrow t}, D\right) +\lambda_{mel} \mathcal{L}_{\mathrm{mel}}\left(G_{s \rightarrow t}\right),
\end{aligned}
\end{equation}
where, $\lambda_{cd}$ is the coefficient of cross-domain consistency loss, $\lambda_{fm}$ is the coefficient of feature matching Loss, and $\lambda_{mel}$ is the coefficient of mel-spectrogram loss, which are $10^3$, 2 and 45  respectively.

\section{Experiments}\label{Experiments}
\subsection{Dataset}
For comparison with other models, we use the AISHELL3 dataset \cite{shi2020aishell} and CSMSC dataset\cite{baker} as source datasets, respectively. The AISHELL3 is a large-scale and high-fidelity multi-speaker Mandarin speech corpus, which contains roughly 85 hours of emotion-neutral recordings spoken by 218 native Chinese Mandarin speakers. It is worth mentioning that the age centers on young adults around 20 years old among all speakers. The AISHELL3 is divided into two parts, the training set and the test set, and the next experiments will be carried out according to this division. The CSMSC is a single-speaker Chinese speech corpus, which contains 10,000 recordings totaling about 12 hours. We randomly assign 200 items as the test set and the rest as the training set. The internal dataset VXI-Children is used as the target dataset, which is a small dataset containing only 20 short recordings from a 6-year-old boy. We randomly assign 10 samples as the training set and the other 10 samples as the test set. To evaluate audio quality, we performe mean opinion score (MOS) scoring with 10 raters. The MOS scores were recorded with 95\% confidence intervals (CI).

\subsection{Results}
First, we train MelGAN and HiFi-GAN V1 with the source dataset, and all the models are trained until 2.5M steps. Next, we use the target dataset for transfer learning. In this process, the traditional GAN transfer learning method based on Equation 1 and the transfer learning method based on cross-domain consistency loss are used respectively. The number of steps for transfer learning of all models is 10K. To verify the transfer effect of our proposed model, we train a source model with strong robustness (the source dataset is AISHELL3) and a source model with poor robustness (the source dataset is CSMSC), respectively.

Tables 1 and 2 display the results for the source data of AISHELL3. Table 1 shows the MOS score of the source model in the AISHELL3 test set. Table 2 compares the MOS scores of the non-transfer method and the two transfer methods on the target set. From the results of test set, we know the MOS score without transfer learning is always the lowest, which may be due to the large difference in the distribution of source and target data sets. In transfer learning, the method proposed in this paper is obviously better than the traditional GAN transfer learning method.

\begin{table}[th]
  \caption{MOS score comparison of the vocoders trained on the source dataset (AISHELL3).}
  \label{tab:mos1}
  \centering
  \begin{tabular}{cccc}
  \toprule
       Model & Ground Truth & MelGAN & HiFi-GAN  \\
      \midrule
       MOS  & 4.52 ($\pm$ 0.05) & 3.68 ($\pm$ 0.09) & 4.32 ($\pm$ 0.06) \\
    \bottomrule
\end{tabular}
\end{table}

\begin{table}[th]
  \caption{MOS score comparison of vocoders obtained by transfer learning from the target dataset. The source dataset is AISHELL3}
  \label{tab:mos2}
  \centering
\begin{tabular}{cccc}
  \toprule
Model  & Fine Tune   & MOS(train) & MOS(test) \\
      \midrule
\multirow{3}{*}{MelGAN}   & Non  & 3.35 ($\pm$0.08)       & 3.34 ($\pm$0.08) \\
                ~          & Traditional & 3.70 ($\pm$0.09)        & 3.12 ($\pm$0.09)      \\
                 ~         & Ours & 3.65 ($\pm$0.06)       & 3.67 ($\pm$0.06)     \\
      \midrule
\multirow{3}{*}{HiFi-GAN} & Non  & 4.12 ($\pm$0.07)       & 4.13 ($\pm$0.06)      \\
                  ~        & Traditional & 4.30 ($\pm$0.08)       & 3.95 ($\pm$0.08)      \\
                   ~       & Ours & 4.32 ($\pm$0.06)       & 4.32 ($\pm$0.06)     \\
    \bottomrule
\end{tabular}
\end{table}

Tables 3 and 4 are the results for the source data of CSMSC from which we find similar results as above. At the same time, we find that the model can achieve better results through transfer learning even in the source model with poor robustness.

\begin{table}[th]
  \caption{MOS score comparison of the vocoders trained on the source dataset (CSMSC).}
  \label{tab:mos3}
  \centering
  \begin{tabular}{cccc}
  \toprule
       Model & Ground Truth & MelGAN & HiFi-GAN  \\
      \midrule
       MOS  & 4.69 ($\pm$ 0.04) & 3.80 ($\pm$ 0.08) & 4.42 ($\pm$ 0.05) \\
    \bottomrule
\end{tabular}
\end{table}

\begin{table}[th]
  \caption{MOS score comparison of vocoders obtained by transfer learning from the target dataset. The source dataset is CSMSC}
  \label{tab:mos4}
  \centering
\begin{tabular}{cccc}
  \toprule
Model  & Fine Tune   & MOS(train) & MOS(test) \\
      \midrule
\multirow{3}{*}{MelGAN}   & Non  & 2.32 ($\pm$0.09)       & 2.35 ($\pm$0.09) \\
                ~          & Traditional & 3.57 ($\pm$0.09)        & 3.12 ($\pm$0.08)      \\
                 ~         & Ours & 3.75 ($\pm$0.05)       & 3.67 ($\pm$0.05)     \\
      \midrule
\multirow{3}{*}{HiFi-GAN} & Non  & 2.25 ($\pm$0.09)       & 2.28 ($\pm$0.09)      \\
                  ~        & Traditional & 4.30 ($\pm$0.06)       & 3.86 ($\pm$0.08)      \\
                   ~       & Ours & 4.42 ($\pm$0.05)       & 4.38 ($\pm$0.05)     \\
    \bottomrule
\end{tabular}
\end{table}

\subsection{End-to-End Custom Voice}
We conducte an additional experiment to examine the effectiveness of the proposed models when applied to an end-to-end custom voice pipeline, which consists of custom acoustic and custom vocoder modules. We herein use AdaSpeech\cite{yan2021adaspeech} as the custom acoustic model, and utilize transfer-learned HiFi-GAN (CSMSC) as custom vocoder. Due to the mismatch between the mel spectrogram predicted by the acoustic model and the mel spectrogram input by the vocoder during the training phase, the method introduced in HiFi-GAN\cite{kong2020hifi} is used to fine-tune the trained custom vockder. The MOS scores are listed in Table 5. We conclude that a natural custom voice system can be constructed with only 10 recordings by combining the AdaSpeech and the fine-tuned AdaHiFi-GAN.

\begin{table}[th]
  \caption{Quality comparison for end-to-end custom voice.}
  \label{tab:mos5}
  \centering
  \begin{tabular}{cc}
  \toprule
       Model & MOS \\
      \midrule
       Ground Truth & 4.65 ($\pm$ 0.06) \\
       AdaSpeech+AdaHiFi-GAN(w/o fine-tuning) & 3.78 ($\pm$ 0.07) \\
       AdaSpeech+AdaHiFi-GAN(fine-tuned) & 4.17 ($\pm$ 0.06) \\
    \bottomrule
\end{tabular}
\end{table}

\section{Conclusions}\label{Conclusion}
In this paper, we propose an adaptive vocoder for custom voice, a novel direction compared with robust vocoders. The adaptive vocoder mainly uses a cross-domain consistency loss to solve the overfitting problem encountered by the GAN-based neural vocoder in the transfer learning of few-shot (less than 10) scenes. The empirical results show that a high-quality custom voice system can be built by combining the adaptive acoustic model with the adaptive vocoder. In the future, it can be considered to apply the cross-domain consistency loss to GAN-based voice conversion systems such as StarGAN-VC. In this way, the transformation of speaker, emotion and other characteristics can be completed with only a small amount of data.

\newpage

\bibliographystyle{IEEEtran}

\bibliography{template}

\end{document}